\documentclass[a4paper]{jpconf}
\usepackage{graphicx}
\usepackage{latexsym}		
\usepackage{amsmath}
\usepackage{amssymb}
\newcommand{\BM}[1]{{\mbox{\boldmath $#1$}}}
\newcommand{\hook}{\raisebox{-0.35ex}{\makebox[0.6em][r]
{\scriptsize $-$}}\hspace{-0.15em}\raisebox{0.25ex}{\makebox[0.4em][l]{\tiny
 $|$}}}
\newcommand{\st}{spacetime}
\newcommand{\be}{\begin{equation}}
\newcommand{\ee}{\end{equation}}
\newcommand{\ba}{\begin{eqnarray}}
\newcommand{\ea}{\end{eqnarray}}

\newcommand{\hh}{\, ,\hspace{0.5cm}}
\newcommand{\hhh}{\, ,\hspace{0.2cm}}

\newcommand{\eq}[1]{Eq.(\ref{#1})}
\newcommand{\n}[1]{\label{#1}}

\newcommand{\ve}{\varepsilon}

\newcommand{\jCQG}[3]{ #2 {\em Class. Quantum Grav.\ }{\bf #1} #3}
\newcommand{\jJMP}[3]{ #2 {\em J. Math. Phys.\ }{\bf #1} #3}
\newcommand{\jPRD}[3]{ #2 {{\em Phys. Rev.}\ } D\ {\bf #1} #3}

\newcommand{\jPLA}[3]{ #2 {{\em Phys. Lett.}\ } A\ {\bf #1} #3}
\newcommand{\jPLB}[3]{ #2 {{\em Phys. Lett.}\ } B\ {\bf #1} #3}
\newcommand{\jPRL}[3]{ #2 {\em Phys. Rev. Lett.\ }{\bf #1} #3}
\newcommand{\jGRG}[3]{ #2 {\em Gen. Rel. Grav.\ }{\bf #1} #3}

\newcommand{\jPRSA}[3]{ #2 {{\em Proc. R. Soc.}\ } A\ {\bf #1} #3}
\newcommand{\jAM}[3]{ #2 {\em Ann. Math.} {\bf #1} #3}
\newcommand{\jIJTP}[3]{ #2 {\em Int. J. Theor. Phys.\ }{\bf #1} #3}
\newcommand{\jTEN}[3]{ #2 {\em Tensor\ }{\bf #1} #3}

\newcommand{\jIJMPA}[3]{ #2 {{\em Int.\ J.\ Mod.\ Phys.}\  } A\ {\bf #1} #3}

\newcommand{\jPTPS}[3]{ #2 {\em Prog. Theor. Phys. Suppl.\ }{\bf #1} #3} 
\newcommand{\jJPA}[3]{ #2 {{\em Journ. Phys.}\ } A\ {\bf #1} #3}
\newcommand{\jJHEP}[3]{ #2 {{\em JHEP}\ }{\bf #1} #3}
\newcommand{\jJGP}[3]{ #2 {{\em J.Geom.Phys.}\ }{\bf #1} #3}

\begin{document}
\title{Hidden Symmetries and Black Holes}

\author{Valeri P. Frolov}

\address{Theoretical Physics Institute, University of Alberta, Edmonton,
Alberta, Canada T6G 2G7}

\ead{frolov@phys.ualberta.ca}

\begin{abstract}
The paper contains a brief review of recent results on hidden
symmetries in higher dimensional black hole spacetimes. We show how
the existence of a principal CKY tensor (that is a closed conformal 
Killing-Yano 2-form) allows one to generate a `tower' of Killing-Yano
and Killing tensors responsible for hidden symmetries. These
symmetries imply complete integrability of geodesic equations and the
complete separation of variables in the Hamilton-Jacobi,
Klein-Gordon, Dirac and gravitational perturbation equations in the
general Kerr-NUT-(A)dS metrics. Equations of the parallel transport of
frames along geodesics in these spacetimes are also integrable.
\end{abstract}

\section{Introduction} 

There are several reasons why a subject of higher dimensional black
holes becomes so popular recently.   The idea that the spacetime can
have more than four dimensions is very old. Kaluza and Klein used
this idea about 80 years ago in their attempts to unify
electromagnetism with gravity. The modern superstring theory is
consistent (free of conformal anomalies) only if the spacetime  has a
fixed number (10) of dimensions. Usually it is assumed that extra
dimensions are compactified. The natural size of compactification in
the string theory is of order of the Planckian scale. In recently
proposed models with large extra-dimensions it is also assumed that
the spacetime has more than 3 spatial dimensions. A new feature is
that the size of the extra dimensions  can be much larger than the
Planckian size, $10^{-33}$cm (up to 0.1mm) . In order to escape
contradictions with observations it is usually assumed that the
matter and fields (except the gravitational one) are confined to a
four-dimensional brane representing our world, while the gravity can
propagate in the bulk. In so called ADD models \cite{ADD1,ADD2} the
extra dimensions are flat. In the Randall-Sundrum  models
\cite{RS1,RS2} the bulk 5D spacetime is curved and it has
anti-deSitter asymptotics. Black holes in the string theory and in
the models with larger extra dimensions play an important role
serving as probes of extra dimension. Study of higher dimensional
black holes is a very important problem of the modern theoretical and
mathematical physics.

One of the main features of the models with large extra
dimensions is a prediction that gravity becomes strong at small
distances. This conclusion implies that for the particle collision
with the energy of the order of TeV the gravitational channel would
be as important as the electroweak channel of the interaction. Under
these conditions two qualitatively new effects are possible: (1) bulk
emission of gravitons, and (2) mini-black-hole production. These
effects have been widely discussed in the connection with the expected
new data at the Large Hadronic Collider (LHC) (see e.g. \cite{Gid}
and references therein). Hawking radiation produced by such mini black
holes has several observable features which allows one, in principle,
to single out such events in observations \cite{Kanti1,Kanti2,park}.

When the gravitational radius of a black hole is much smaller than
the size of extra dimensions and the tension of the brane is
neglected one can consider a the black hole as an isolated one. Such
a black hole is described by a solution of the higher dimensional
Einstein equations which is asymptotically flat or has (A)dS
asymptotics. It was shown that besides `standard' black holes with a 
spherical topology of the horizon, in higher dimensions there exist
black objects with a different horizon topology. Black rings
\cite{BR} and black saturns \cite{BS} are examples of such solutions
in the 5D spacetime. For a review of higher dimensional black objects
and their properties  see, e.g., \cite{Em:2008}. 

The most general known  higher dimensional black hole solution of the
Einstein equations with the spherical topology of the horizon is a 
{\em Kerr-NUT-(A)dS metric} which was discovered recently
\cite{Pope}.  The metric belongs to the special algebraic type {\bf
D} \cite{Riem} of the higher-dimensional algebraic classification 
\cite{C1,C2,C3}. This makes these metrics quite different from the
black ring and black saturn type solutions, which are of type $I_i$
\cite{C3}. In this paper we review properties of the higher
dimensional black hole solutions and, especially, their hidden
symmetries. We demonstrate that the properties of the higher
dimensional black holes, in many aspects are similar to the
properties of the four dimensional black holes.

An isolated stationary black hole in 4-dimensional asymptotically
flat spacetime is uniquely specified by two parameters, its mass and
angular momentum. The corresponding Kerr metric possesses a number of
properties, which was called by Chandrasekhar `miraculous'. In
particular the Kerr metric allows the separation of variables in the
geodesic Hamilton-Jacobi equation and a massless field equations.
These properties look `miraculous' since the spacetime symmetries of
the Kerr metric are not sufficient to explain them.   Really, spacetime
symmetries are `responsible' for two integrals of motion, the energy
and the azimuthal component of the angular momentum. This, together
with the conservation of $\BM{p}^2$, gives only 3 integrals of
motion.  Carter \cite{Car:68a,Car:68c} constructed the fourth
required integral of motion, which is quadratic in momentum and is
connected with the Killing tensor \cite{WP}. Penrose and Floyd
\cite{Penrose} showed that this Killing tensor is a `square' of an
antisymmetric Killing-Yano tensor \cite{Yano}. 

In many aspects a Killing-Yano tensor is more fundamental than a
Killing tensor. Namely, its `square'   is always Killing tensor, but
the opposite is not generally true (see, e.g., \cite{Ferrando}).   It
was shown by Collinson \cite{Coll74} a $4$--dimensional vacuum
spacetime which admits a non--degenerate Killing-Yano tensor  is of
the type $\bf D$.  All the vacuum type $\bf D$ solutions were
obtained by Kinnersley \cite{Kinn}. Demianski and Francaviglia
\cite{DeFr} showed that in the absence  of the  acceleration these
solutions admit Killing and Killing-Yano tensors.  It should be  also
mentioned that if a spacetime admits a non--degenerate Killing-Yano
tensor it always has at least one Killing vector  \cite{Dietz}.

Separability of massless and massive field equations (including the
gravitational perturbations) in the 4D Kerr metric 
\cite{Teuk_a,Unruh,Teuk_b,Chandrasekhar,Page76} is a direct
consequence of the hidden symmetries. The separability property plays
a key role in study of properties of rotating black holes, including
the proof of their stability and the calculation of Hawking
radiation. 

In 2003 is was discovered that five-dimensional Myers-Perry black
holes possess a Killing tensor \cite{FS1,FS2}. This makes geodesic
equations completely integrable (see e.g \cite{GF} where the
cross-section of particles and photons for 5D black holes was
studied.) After this a lot of attempts were made to study the hidden
symmetries of the higher dimensional black holes. In the most of these
publications an additional assumption was made. Namely special
conditions were imposed on the rotation parameters of a black hole
which enhance the symmetry of these solutions. In 2007 it was
discovered that Myers-Perry metrics with arbitrary rotation
parameters, as well as a general Kerr-NUT-(A)dS metrics possesses a
closed conformal Killing-Yano 2-form \cite{FK,KF}. This result was used to
demonstrate that the general high dimensional black hole solutions in
many aspects are similar to their four dimensional `cousin'.
This paper contains a brief review of the recent development of the
theory of higher dimensional black holes and hidden symmetries (see
also \cite{Fro,CQG,TH}).

\begin{figure}[h]
\begin{center}
\includegraphics[width=14pc]{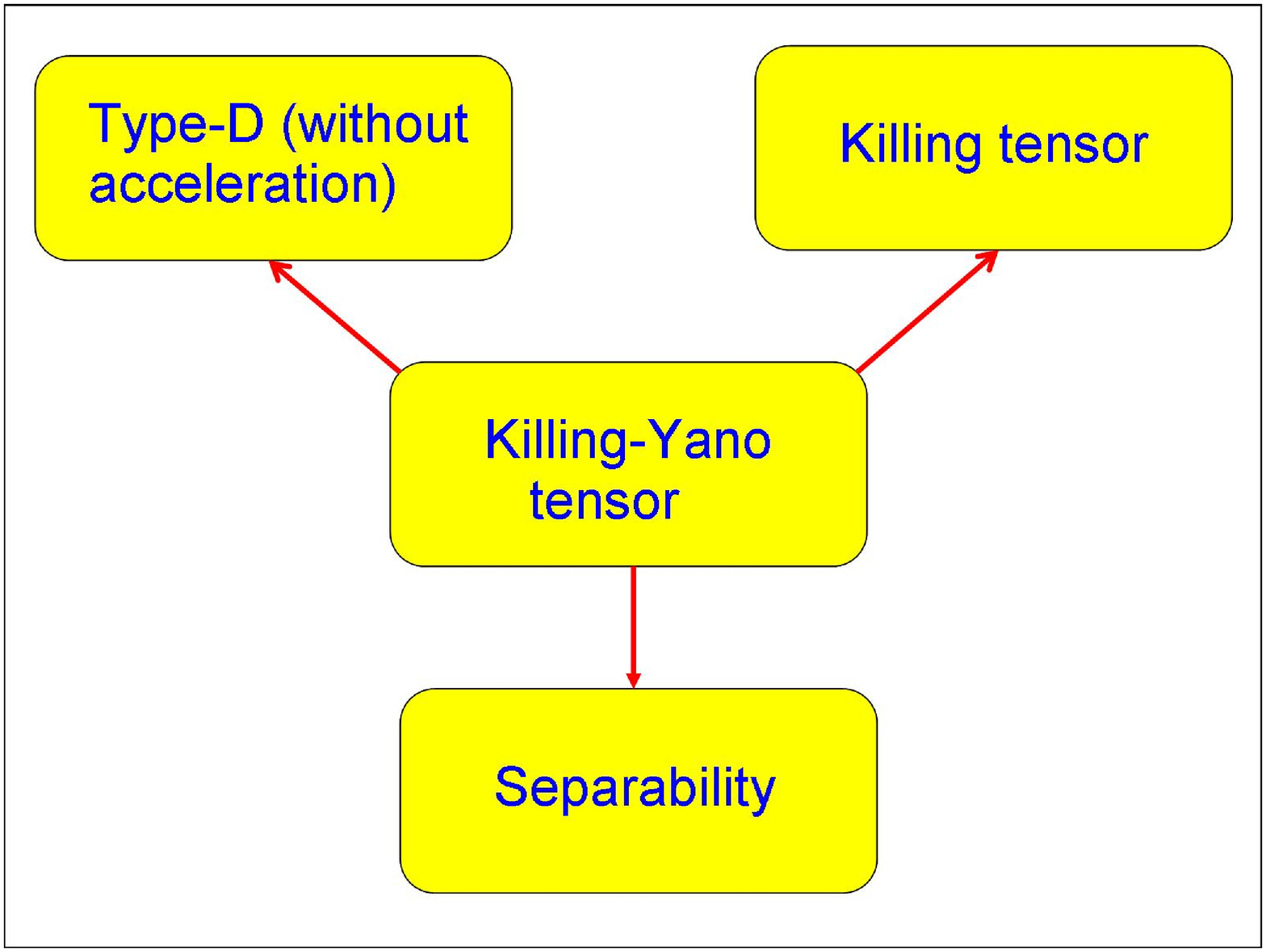}\hspace{2pc}%
\end{center}
\caption{\n{gen}Killing-Yano tensors in 4D \st\ and their properties.}
\end{figure}

\section{Hidden symmetries}

The concept of symmetries is one of the most powerful tools of modern
theoretical physics. Noether's theorem relates continuous symmetries 
to conservation laws. The most fundamental of them are connected with
the symmetries of the background spacetime. A curved spacetime 
possesses a symmetry if there exists a diffeomorphism
preserving the geometry. Consider  a one-parameter family of continuous 
transformations
\be\n{tran}
x^{a }\to \bar{x}^{a }=F^{a }(x,t)\, .
\ee
Denote by $\BM{\xi}$ a vector field generating these transformations
\be
\xi^{a }=\left. {dF^{a }\over dt}\right|_{t=0}\, .
\ee
Invariance of the metric $\BM{g}$ under the transformations \eq{tran}
implies that
\be\n{kil1}
{\cal L}_{\BM{\xi}}g_{a b }=0\, ,
\ee
where ${\cal L}_{\BM{\xi}}$ is the Lie derivative. A generator
$\BM{\xi}$ of the continuous symmetry transformation (isometry) is
called a {\em Killing vector}. The equation \eq{kil1} can be identically
rewritten in the form
\be\n{kil2}
\xi_{(a ;b )}=0\, .
\ee

In the presence of the symmetry generated by the Killing vector
$\BM{\xi}$  geodesic equations possess an
integral of motion
\be\n{Pk}
P_{\xi}=\xi_{a }u^{a }\, ,
\ee
where $u^{a }=dx^{a }/d\lambda$ is a tangent vector to a geodesic
and $\lambda$ is an affine parameter. Really,
\be
{dP_{\xi}\over d\lambda}=u^{b }u^{a }\xi_{a ;b }+u^{b }u^{a }_{\
;b }\xi_{a }=0\, .
\ee
The first term in the right hand side vanishes because of the Killing
equation \eq{kil2}, while the second one vanishes because of the
geodesic equation of motion $u^{b }u^{a }_{\ ;b }=0$.

\begin{figure}[h]
\begin{center}
\includegraphics[width=14pc]{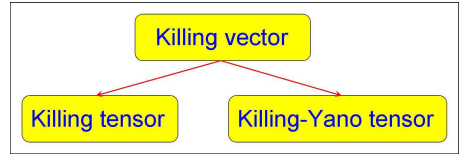}\hspace{2pc}%
\end{center}
\caption{\n{gen}Possible generalizations of the Killing vector.}
\end{figure}

A Killing tensor $K_{a _1\ldots a _k}$ is a natural {\em symmetric
generalization} of the Killing vector. Let us assume that for any
geodesic with a tangent vector $u^{a }$ the following object
\be
P_{K}=K_{a _1\ldots a _k}u^{a _1}\ldots u^{a _k}\, 
\ee 
is concerved
\be
{dP_{K}\over d\lambda}=0\, .
\ee
For a geodesic motion
\be\n{KK}
{dP_{K}\over d\lambda}\equiv u^{b }u^{a _1}\ldots u^{a _k} K_{a
_1\ldots a _k;b }
\ee
Since this relation is valid for an arbitrary $\BM{u}$ one has
\be\n{KT}
K_{(a _1\ldots a _k;b )}=0\, .
\ee
A symmetric tensor $K_{(a _1\ldots a _k)}$ obeying the relation
\eq{KT} is called a {\em Killing tensor}.

A Killing-Yano tensor $f_{a _1\ldots a _k}$ is an  {\em antisymmetric
generalization} of the Killing vector. Let us assume that for any
geodesic with a tangent vector $u^{a }$ the following object
\be
P_{a _1\ldots a _{k-1}}=f_{a _1\ldots a _k}u^{a _k}\, 
\ee 
is parallel propagated along the geodesic
\be
u^{b }P_{a _1\ldots a _{k-1};b }=0\, .
\ee
Using the geodesic equation one obtains
\be
f_{a _1\ldots a _k;b }u^{a _k}u^{b }=0\, .
\ee
Since this relation is valid for an arbitrary $u^{a }$ it implies
\be
f_{a _1\ldots(a _k;b )}=0\, .
\ee
A skewsymmetric tensor $f_{a _1\ldots a _k}$ obeying this relation is
called a {\em Killing-Yano tensor}.

It is easy to show that if $f_{a _1\ldots a _k}$ is a Killing-Yano
tensor then
\be\n{KYKT}
K_{a b }=f_{a a _1\ldots a _{k-1}}f^{\ \ a _1\ldots a _{k-1}}_{b }
\ee
is a Killing tensor.

An important generalization of the symmetry \eq{kil1} is a {\em
conformal symmetry}. A conformal Killing vector $\BM{\xi}$ generating such a
transformation obeys the equation
\be\n{ckil1}
{\cal L}_{\BM{\xi}}g_{a b }=\beta(x) g_{a b }\, ,
\ee
or, equivalently
\be\n{ckil2}
\xi_{(a;b)}=\tilde{\xi}g_{ab}\hh \tilde{\xi}=D^{-1}\xi^{b}_{\ \ ;b}\, .
\ee
The corresponding expression \eq{Pk} is conserved for null geodesics. The
conformal generalizations of the Killing and Killing-Yano tensors are
defined as follows.
A {\em symmetric} tensor $K_{a_1a_2 \ldots a_p}=
K_{(a_1a_2 \ldots a_p)}$ is called a
{conformal Killing tensor} if it obeys the equation
\be\n{CK}
K_{(a_1a_2 \ldots a_p;b)}=g_{b (a_1}\tilde{K}_{a_2 \ldots
a_p)}
\, .
\ee
It is easy to show that a symmetrized tensor product of two conformal Killing
tensors is again a conformal Killing tensor. Similarly, a tensor
product of two Killing tensors is again a  Killing tensor. A
(conformal) Killing tensor is called {\em reducible} if it can be
written as a linear combination of tensor products of lower rank
(conformal) Killing tensor. If a Killing tensor of rank $q$ is
reducible, the corresponding conserved quantity for a geodesic motion,
which is a polynomial of rank $q$ in momentum, can be written as a
linear combination of products of conserved quantities of lower than
$q$ powers in momentum. It means that a reducible Killing tensor does
not generate any new {\em independent} conserved quantities.

An {\em antisymmetric} generalization of the conformal Killing vector is known
as a {\em conformal Killing-Yano tensor}. An antisymmetric tensor 
$h_{a_1a_2 \ldots a_p}=h_{[a_1a_2 \ldots a_p]}$ is called
a {\em conformal Killing-Yano tensor} (or, briefly, CKY tensor) if it
obeys the following equation \cite{Tachibana}
\be\n{CKY}
\nabla_{(a_1}h_{a_2)a_3 \ldots a_{p+1}}=
g_{a_1a_2}\tilde{h}_{a_3 \ldots a_{p+1}}-
(p-1)g_{[a_3(a_1}\tilde{h}_{a_2) \ldots a_{p+1}]}\, .
\ee 
By tracing the both sides of this equation one obtains the following
expression for $\tilde{\BM{h}}$
\be
\tilde{h}_{a_2 a_3 \ldots a_{p}}={1\over D-p+1}\nabla^{a_1}h_{a_1a_2
\ldots a_p}\, .
\ee

\section{Conformal Killing-Yano tensors}

Let us discuss properties of the conformal Killing-Yano (CKY) tensors in
more details. 
The CKY tensors are forms and operations with them are greatly
simplified if one uses the "language" of differential forms. 
We just remind some of the relations we use in the present paper.
If $\BM{\alpha}_p$ and $\BM{\beta}_q$ are $p$- and $q$-forms,
respectively, the external derivative ($d$) of their external product
($\wedge$) obeys a relation
\be\n{dd}
d(\BM{\alpha}_p\wedge \BM{\beta}_q)=
d\BM{\alpha}_p\wedge \BM{\beta}_q+
(-1)^p\BM{\alpha}_p\wedge d\BM{\beta}_q\, .
\ee
A
Hodge dual $*\BM{\alpha}_p$ of the $p$-form $\BM{\alpha}_p$ is
$(D-p)$-form defined as
\be
* \BM{\alpha}_p \Leftrightarrow 
(* \alpha)_{a_1\ldots a_{D-p}}={1\over p!} 
\alpha^{b_1\ldots b_p}e_{b_1\ldots b_p a_1\ldots a_{D-p}}\, ,
\ee
where $e_{a_1 \ldots a_D}$ is a totally anti-symmetric tensor.
The exterior {\em co-derivative}
$\delta $ is defined as follows
\be
\delta \BM{\alpha}_p=(-1)^p\epsilon_p *d*\BM{\alpha}_p\hhh
\epsilon_p=(-1)^{p(D-p)}{\mbox{det}(g)\over |\mbox{det}(g)|}\, .
\ee
One also has $* *\BM{\alpha}_p=\epsilon_p\BM{\alpha}_p$.

If $\{ \BM{e}_a\}$ is a basis of vectors, then dual basis of 1-forms
$\BM{\omega}^a$ is defined by the relations
$\BM{\omega}^a(\BM{e}_b)=\delta^a_b$. We denote
$\eta_{ab}=g(\BM{e}_a,\BM{e}_b)$ and by $\eta^{ab}$ the inverse
matrix. Then the operations with the indices enumerating the basic
vectors and forms are performed by using these matrices. In
particular, $\BM{e}^a=\eta^{ab}\BM{e}_b$, and so on. We denote a
covariant derivative along the vector $\BM{e}_a$ by
$\nabla_a=\nabla_{\BM{e}_a}$. One has
\be
d=\BM{\omega}^a\wedge \nabla_a\hhh
\delta=-\BM{e}^a\hook \nabla_a\, .
\ee
In the tensor notations the `hook' operator applied to a $p$-form
$\BM{\alpha}_p$ corresponds to a contraction 
\be
\BM{X}\hook \BM{\alpha}_p \Leftrightarrow X^{a_1} \alpha_{a_1 a_2
\ldots a_p}\, .
\ee
For a given vector $\BM{X}$ one  defines $\BM{X}^{\flat}$   as a
corresponding 1-form with the components
$(X^{\flat})_{a}=g_{ab}X^{b}$. 
In particular, one has $\eta^{ab}(\BM{e}_b)^{\flat}=\BM{\omega}^a$.

The definition \eq{CKY} of the CKY tensor $\BM{h}$ (which is a
$p$-form)  is equivalent to the following equation (see e.g.
\cite{BCK,kress})  
\be\n{CKYf}
\nabla_\BM{X} \BM{h}={1\over p+1} \BM{X}\hook d\BM{h}
-{1\over D-p+1}\BM{X}^{\flat}\wedge
\delta \BM{h}\, .
\ee 
If $\delta \BM{h}=0$ this is an equation for the Killing-Yano tensor.

The CKY tensors possess the following properties:
\begin{enumerate}
\item If $\BM{\alpha}$ is a CKY tensor then $*\BM{\alpha}$ is also
a CKY tensor;
\item If $\BM{\alpha}$ is a closed CKY tensor ($d\BM{\alpha}=0$) then
$*\BM{\alpha}$ is  a Killing-Yano (KY) tensor;
\item If $\BM{\alpha}$ and $\BM{\beta}$ are closed CKY tensors then 
$\BM{\alpha}\wedge \BM{\beta}$ is also a closed CKY tensor.
\end{enumerate}

The first two properties can be proved by using a relation
\be
\BM{X}\hook *\BM{\omega}=*(\BM{\omega}\wedge \BM{X}^{\flat})\, .
\ee
Applying this relation to \eq{CKYf} one has
\be
\nabla_\BM{X}(* \BM{h})={1\over p_*+1} \BM{X}\hook d(*\BM{h})
-{1\over D-p_*+1}\BM{X}^{\flat}\wedge
\delta(*\BM{h})\hhh p_*=D-p\, .
\ee
It means that a Hodge dual $*\BM{h}$ of a CKY tensor $\BM{h}$ is again
a CKY tensor. Moreover, if the CKY is closed, $d\BM{h}=0$, then its
dual $(D-p)$-form $f=*\BM{h}$ is a Killing-Yano tensor. The proof of
the third property can be found in \cite{Fro,KKPF}.
This property means that the closed CKY tensors form an algebra.

\section{Principal CKY tensor and Killing-Yano tower} 

Consider a $D$-dimensional \st. We write \be D=2n+\ve\, , \ee where
$\ve=1$ for the odd number of dimensions and  and $\ve=0$ for the
even  number. Consider a 2-form $h_{ab}$ which is a CKY tensor. We
assume that it is closed, $d\BM{h}=0$, and non-degenerate, that is it
has a matrix rank $2n$. We call such a tensor a {\em principal CKY
tensor}. The principal CKY tensor obeys the dollowing equation
\be\n{PCKY}
\nabla_{X}\BM{h}=\BM{X}^{\flat}\wedge \BM{\xi}^\flat\hh
\xi_b=\frac{1}{D-1}\nabla_dh^{d}_{\ b}\,,
\ee 
where $\BM{X}$ is an arbitrary vector field. 

Starting with the principal CKY tensor one can construct a  a
Killing-Yano tower of tensors \cite{KKPF,PKVK}. The following diagram
illustrates this construction.

\begin{figure}[h]
\begin{center}
\includegraphics[width=32pc]{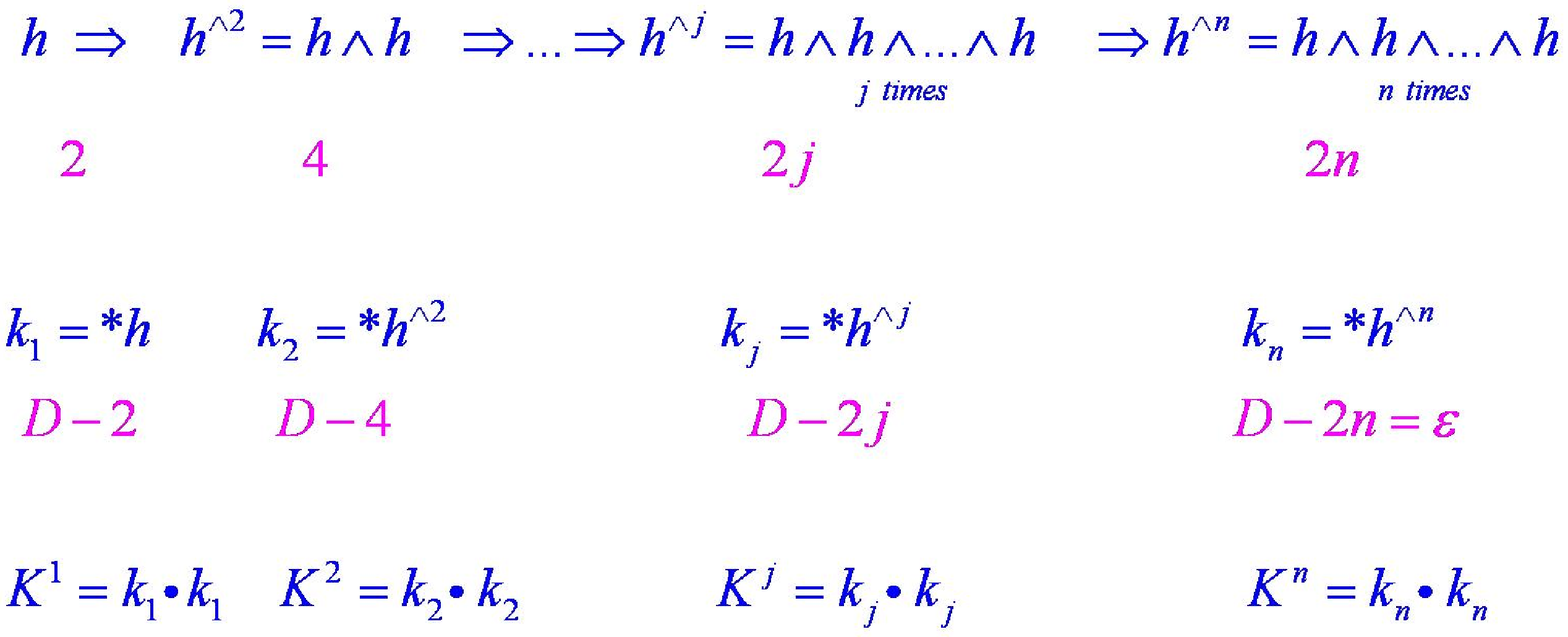}\hspace{2pc}%
\end{center}
\caption{\n{KYT}Killing-Yano tower}
\end{figure}

The first row of this diagram contains external powers
$\BM{h}^{\wedge k}$ of the principal CKY tensor, which again are
closed CKY tensors. Taking the Hodge dual of these tensors one
obtains a set of Killing-Yano tensors $\BM{k}_i$. `Squares' of
$\BM{k}_i$ give a set of second rank Killing tensors $\BM{K}^i$. For
even $D$ the last column can be omitted since $\BM{h}^{\wedge n}$ is
proportional to a totally antisymmetric tensor, and hence it does not
produce a non-trivial Killing tensor. For the odd case, the Killing
tensor $\BM{K}^n$ is a product of two Killing vectors and hence it is
reducible. The first $n-1$ Killing tensors are irreducible. The metric
itself is also a Killing tensor, $\BM{K}^0=\BM{g}$. We call the
Killing-Yano tower {\em extended} if the metric is included into it.
Thus the extended Killing-Yano tower allows one to construct $n$
independent quadratic in momenta conserved quantities for a geodesic
motion.

Besides a tower of the rank 2 Killing tensors, the principal CKY
tensor generates a set of the Killing vectors. To demonstrate this let
us notice that
for a CKY tensor $\BM{h}$ of rank-$2$ the vector
\be\n{prime}
\xi^{(0)a}={1\over D-1}\nabla_b h^{ab}
\ee
obeys the following equation \cite{Tachibana,jez}
\be\label{Tachibana}
\xi^{(0)}_{\ (a;b)}=-{1\over D-2}R_{c (a}h_{b)}^{\,\,\,\,c}\,
.
\ee
Thus, in an Einstein space, that is when
$R_{ab}=\Lambda g_{ab}$, $\BM{\xi}^{(0)}$ is the Killing vector.
It can be shown that even if the Einstein equations with the
cosmological constant are not imposed the vector $\xi^{(0)a}$
constructed for the principal CKY tensor is always the Killing
vector\cite{KFK}. We call it a {\em primary Killing vector}. Acting
by the Killing tensors $\BM{K}^i$ on the primary Killing vector one
obtains a set of the independent commuting Killing
vectors\cite{KKPF,KFK,hoy1}
\be\n{second}
\xi^{i}_a=K^{i}_{ab}\xi^{(0)b}\, .
\ee
There Killing vectors (with an additional Killing vector $\BM{k}_n$
for $\ve=1$) give $n+\ve$ conserved first order in momentum
quantities for the geodesic motion. These conserved quantities
together with $n-$  conserved quantities connected with the Killing
tensors (including the metric $\BM{g}$) form a set of $D=2n+\ve$
conserved quantities.

\section{Darboux basis}

Let us consider an eigenvalue
problem for a conformal Killing tensor $\BM{H}$ associated with $\BM{h}$ 
\be \n{HH}
H_{ab}\equiv h_{ac}h_b^{\ \,c}\,.
\ee
It is easy to show that in the Euclidean domain
its eigenvalues $x^2$, 
\be\n{HHH}
H^a_{\ b}v^{b}=x^2 v^a\, ,
\ee 
are real and non-negative. Using a modified Gram--Schmidt procedure it
is possible to show that there exists such an orthonormal basis in
which the operator $\BM{h}$ has the following structure:
\be\n{dar}
\mbox{diag}(\BM{0},\BM{\Lambda}_1,\ldots,\BM{\Lambda}_p)\, ,
\ee
where $\BM{\Lambda}_i$ are matrices of the form
\be
\Lambda_i=\left(   
\begin{array}{cc}
0 & -x_i \BM{I}\\
x_i \BM{I}& 0
\end{array}
\right)\, ,
\ee
$\BM{0}$ is a zero matrix and $\BM{I}$ are unit matrices. For a
non-degenerate principal CKY tensor $\BM{h}$ the eigenspaces
corresponding to the eigenvalues $x_i$ are two dimensional ({\em
Darboux subspaces}). Since the matrix rank of $\BM{h}$ is $2n$ the
eigenvalue $0$ is present only for $\ve=1$, and the corresponding
eigenspace in this case is one dimensional.  We assume also that
$\BM{h}$ is non-degenerate, that is $p=n$ and the eigenspaces of
$\BM{H}$ are two-dimensional so that the matrices $\BM{\Lambda}$ has the
form
\be
\Lambda_i=\left(   
\begin{array}{cc}
0 & -x_i \\
x_i & 0
\end{array}
\right)\, .
\ee
(See Section~9 for a discussion of a degenerate case.)

We denote  by $\BM{e}_{\hat \mu}$ and $\BM{\tilde e}_{\hat \mu}\equiv
\BM{e}_{\hat n+\hat \mu}$, where $\mu=1,\ldots,n$,  mutually
orthogonal unit vectors in the 2D Darboux space corresponding to the
eigenvalue $x_{\mu}$. For $\ve=1$ we introduce also a unit vector
$\BM{e}_{\hat 0}$ in the subspace corresponding to zero eigenvalue.
The basis of the dual forms we denote by $\BM{\omega}^{\hat \mu}$ and
$\BM{\tilde \omega}^{\hat \mu}\equiv \BM{\omega}^{\hat n+\hat \mu}$
(and $\BM{\omega}^{\hat 0}$ if $\ve=1$). The metric $\BM{g}$ and the
principal CKY tensor $\BM{h}$ in this basis take the form
\ba
\BM{g}\!\!&=&=\sum_{\mu=1}^n 
(\BM{\omega}^{\hat \mu}\BM{\omega}^{\hat \mu}+
\BM{\tilde \omega}^{\hat \mu}\BM{\tilde \omega}^{\hat \mu})
+\ve \BM{\omega}^{\hat 0}\BM{\omega}^{\hat 0}\, ,\n{gab}\\
\BM{h}\!\!&=&\!\!\sum_{\mu=1}^n x_{\mu} 
\BM{\omega}^{\hat \mu}\wedge \BM{\tilde \omega}^{\hat \mu}\,
.\n{hab}
\ea
Starting with the principal CKY tensor $\BM{h}$ written in the {\em
canonical form}, one can find the explicit expressions for the other
associated tensors of the Killing tower. In particular, for the
associated Killing tensors one has
\ba\n{Kdar}
\BM{K}^{(j)} &=&\sum_{\mu=1}^n A^{(j)}_\mu(\BM{\omega}^{\hat
\mu}\BM{\omega}^{\hat \mu}+ \BM{\tilde \omega}^{\hat
\mu}\BM{\tilde \omega}^{\hat \mu})+\ve A^{(j)}\BM{\omega}^{\hat
0}\BM{\omega}^{\hat 0}\,\\
A^{(j)}\!\!&=&\!\!\sum_{\nu_1<\dots<\nu_j}\! 
x_{\nu_1}^2\dots x_{\nu_j}^2\;,\quad
A^{(j)}_\mu=\!\! \sum_{{\nu_1<\dots<\nu_j\atop \nu_i\ne\mu}}\!
   x_{\nu_1}^2\dots x_{\nu_j}^2\, . \n{AAA}
\ea
It should be emphasized that Killing tensors are defined up to an
arbitrary constant factor.  In order to obtain the above expressions
this factors were specially chosen, so that $\BM{K}^{(j)}$ in
\eq{Kdar} contains an extra constant factors with respect to
$\BM{K}^j$ from the diagram \ref{KYT} for the Killing tower.

\section{Canonical form of the metric}

In the presence of the principal CKY tensor, one can use its
independent eigenvalues $x_i$ as $n$ coordinates. Moreover, the
principal CKY tensor generates $n+\ve$ Killing vectors, one primary
$\xi^{(0)}$, \eq{prime}, and $n-1+\ve$ secondary $\xi^{(i)}$, defined
by \eq{second} with $\BM{K}^i$ substituted by $\BM{K}^{(i)}$, ones.
Denote by $\psi_{0}$ and $\psi_{i}$ the corresponding Killing
parameters, 
\be
\xi^{(0)}=\partial_{\psi_0}\hh
\xi^{(i)}=\partial_{\psi_i}\, .
\ee
A set of $n$ {\em essential} (Darboux) coordinates $x_i$ and $n+\ve$ {\em
Killing} coordinates $\{\psi_0,\psi_i\}$ can be used as $D=2n+\ve$
coordinates associated with the principal CKY tensor. We call these
coordinates {\em canonical}. It can be shown that the metric
of a \st, which admits a principal CKY tensor, can be written in the
canonical coordinates in the form\footnote{A similar canonical form of
a $4D$ metric possessing a Killing-Yano tensor was obtained by Carter
\cite{Car}}
\ba\n{can}
\BM{\omega}^{\hat \mu}&=&\frac{1}{\sqrt{Q_\mu}}\,d  x_\mu\hhh 
\BM{\tilde \omega}^{\hat \mu}
=\sqrt{Q_\mu}\,\sum_{k=0}^{n-1}A^{(k)}_\mu\, d \psi_k\hhh
\BM{\omega}^{0} = (-c/A^{(n)})^{1/2}
\sum_{k=0}^nA^{(k)}\BM{d}\psi_k\, ,\\
Q_{\mu}&=&X_{\mu}/U_{\mu}\,,\quad
U_{\mu}=\prod_{\nu\ne\mu}(x_{\nu}^2-x_{\mu}^2)\, .
\ea
This result was first proved in \cite{hoy2} assuming the the following two
conditions
\be
{\cal L}_{\xi}\BM{g}=0\hh {\cal L}_{\xi}\BM{h}=0\, ,
\ee
where $\BM{\xi}$ is defined by \eq{PCKY},
are satisfied. Later it was shown \cite{KFK} that these conditions follow from the
very existence of the principal CKY tensor and the canonical form was
derived in a general case for a \st \ with the principal CKY tensor
\cite{KFK,hoy4}.
Here $A^{(k)}_\mu$ and $A^{(k)}$ are given by \eq{AAA}, and $X_{\mu}$
are functions of a single argument $x_{\mu}$. 
It should be emphasized that in the derivation of this form of the
metric the Einstein equations were not used. In this sense, this is an
{\em off-shell} result. 

By imposing the $D$-dimensional Einstein equations with a
cosmological constant one obtains that the functions $X_{\mu}$ are of
the form\footnote{The components of the curvature tensor for the
metric \eq{can} were calculated in \cite{Riem}.}
\be\n{knds}
X_{\mu}=\sum\limits_{k=\varepsilon}^{n}c_kx_{\mu}^{2k}
-2b_{\mu}x_{\mu}^{1-\varepsilon}
+\frac{\varepsilon c}{x_{\mu}^2}\,.
\ee
`Time' is denoted by $\psi_0$, azimuthal coordinates by $\psi_k$,
${k=1,\dots,m=n-1+\ve}$, and ${x_\mu}$, ${\mu=1,\dots,n}$, stand
for   `radial' and latitude coordinates. The physical metric with
proper signature is recovered when standard radial coordinate
$r=-ix_n$ and new parameter $M=(-i)^{1+\epsilon}b_n$ are introduced. 
The total number of constants which enter the solution is $2n+1$:
$\ve$ constants $c$, $n+1-\ve$ constants $c_k$ and $n$ constants
$b_{\mu}$. The form of the metric is invariant under a 1-parameter 
scaling coordinate transformations, thus a total number of
independent parameters is $D-\ve$.  These parameters are related to
the cosmological constant, mass, angular momenta, and NUT parameters.
One of them may be used to define a scale, while the other $D-1-\ve$
parameters can be made dimensionless. (For more details see
\cite{Pope}.) In the absence of NUT parameters and for vanishing
cosmological constant the metric \eq{can}-\eq{knds} reduces to the
Myers-Perry metric describing an isolated rotating higher dimensional
black hole in an asymptotically flat \st. The existence of the closed
CKY tensor for Myers-Perry \cite{MP} and Kerr-NUT-(A)dS was established first in
\cite{FK,KF}, where it was also demonstrated that the corresponding
principal CKY tensor does not depend on the parameters of the
solution ({\em universality property}).

\section{Separation of variables}

The Hamilton-Jacobi equation for geodesic motion 
\be
{\partial S\over \partial \lambda}+g^{ab}\partial_a S \partial_a S=0\,,
\ee
in the Kerr-NUT-(A)dS spacetime allows a complete separation of
variables \cite{FKK}
\be
S=-w\lambda+\sum_{k=0}^{n+\ve-1}\Psi_k\psi_k+\sum_{\mu=1}^n
S_{\mu}(x_{\mu})\, .
\ee
The functions $S_{\mu}$ obey the first order ordinary differential
equations
\be
{S'_{\mu}}^2={V_{\mu}\over X_{\mu}}-{W_{\mu}^2\over X_{\mu}^2}\, ,
\ee
Here
\be\n{fun}
W_{\mu}=\sum_{k=0}^{n+\ve-1}\Psi_k(-x_{\mu}^2)^{n-1-k}\hhh
V_{\mu}=\sum_{k=0}^{n+\ve-1}\kappa_k(-x_{\mu}^2)^{n-1-k}\, .
\ee
For $\ve=1$ we put $\kappa_n=\Psi_n^2/c$. The parameters $\kappa_k$
and $\Psi_k$  are separation constants. The existence of $D$
independent conserved constants which enter these functions implies a
complete integrability of the geodesic motion equations in the
spacetimes admitting the principal CKY tensor (for more details, see
\cite{KKPF,PKVK}).

Similarly, the massive scalar field equation
\be\n{seq}
\Box \Phi-\mu^2\Phi=0\, ,
\ee
in the Kerr-NUT-(A)dS metric allows a complete separation of variables
\cite{FKK}. Namely, the solution can be decomposed into modes
\be\n{mod}
\Phi=\prod_{\mu=1}^n R_{\mu}(x_{\mu})\prod_{k=0}^{n+\ve-1}e^{i\Psi_k
\psi_k}\, .
\ee
Substitution of \eq{mod} into the equation \eq{seq} results in the
following {\em second order ordinary differential equations} for functions
$R_{\mu}(x_{\mu})$
\be
(X_{\mu}R'_{\mu})'+\ve{X_{\mu}\over
x_{\mu}}R'_{\mu}+\left(V_{\mu}-{W_{\mu}^2\over X_{\mu}}\right)R_{\mu}=0\, .
\ee
Here $W_{\mu}$ and $V_{\mu}$ are given by \eq{fun} and $\kappa_0=-\mu^2$

In \cite{sekr} it was shown that the following
operators 
\ba\n{LLL}
{\hat \xi}_{(k)}&=&-i\xi^{(k)a}\partial_{a}\,,\quad k=0,\ldots,m\, ,\\
{\hat K}_{(j)}&=&-{1\over\sqrt{|g|}}\,
\partial_{a}\Bigl(\sqrt{|g|}K^{(j)\,\! a  b}\partial_{b}\Bigr)
\,,\quad j=0,\ldots,n-1\, ,
\n{KKK}
\ea
determined by a principal CKY tensor, form a complete set of commuting operators 
for the Klein--Gordon equation in the Kerr-NUT-(A)dS background.

Using \eq{KKK} one has ${\hat K}_{(0)}=-\Box$.
Since all the operators \eq{LLL}--\eq{KKK} commute with one another,
their common eigenvalues can be used to specify the modes. It is possible
to show \cite{sekr} that the eigen-vectors of these commuting operators are
the modes \eq{mod} and one has
\be
{\hat \xi}_{(k)}\Phi=\Psi_k\Phi\,,\quad 
{\hat K}_{(j)}\Phi=\kappa_j \Phi\, .
\ee

Later it was shown that the massive Dirac equation in the
Kerr-NUT-(A)dS spacetime also allows the separation of variables
\cite{oy1}. More recently Oota and Yasui \cite{TENSOR} demonstrated
separability of the tensor type gravitational perturbations in a
(generalized) Kerr-NUT-(A)dS spacetime. It was also proved that the
stationary test string equations in the  Kerr-NUT-(A)dS spacetime are
completely integrable \cite{KF_string} .

\section{Parallel transport of frames along geodesics}

One of the additional remarkable properties of the 4D Kerr metric, 
discovered by Marck in 1983 \cite{M1,M2,KM}, is that  the equations
of parallel transport can be  integrated.  This result allows a
generalization: In a higher dimensional \st \ which admits a
principal CKY tensor  equations of a parallel-propagated frame along
a geodesic can be solved explicitly \cite{CFK,KFKC}.

The main idea of this construction is the following. Let $\BM{u}$ be a
tangent vector to a geodesic. Consider a 2-form $\BM{F}$ which is
obtained by projecting $\BM{h}$ onto a subspace $V$ orthogonal to $\BM{u}$
\be
F_{ab}=P_a^c P_b^d h_{cd}=h_{ab}+u_{a}u^{c} h_{cb}+h_{ac}u^{c}u_{b}\,
,
\ee
where ${P_a^b=\delta_a^b+u^bu_a}$ is the projector onto $V$. This form
$\BM{F}$ can be also written as follows
\ba\n{fh}
\BM{F}&=&-\BM{u}\hook(\BM{u}^\flat\wedge \BM{h})=\BM{h}+\BM{u}^{\flat}\wedge \BM{s}\, ,\n{Fop}\\
\BM{s}&=&\BM{u}\hook \BM{h}\hh s_b=u^a h_{ab}\, .\n{ss}
\ea
Let us demonstrate that $\BM{F}$ is parallel propagated along a
geodesic. The definition \eq{PCKY} of the principal CKY tensor implies 
\be
\nabla_{u}\BM{h}=\BM{u}^{\flat}\wedge \BM{\xi}^\flat\, .
\ee
Hence
\be
\nabla_u (\BM{u}^\flat\wedge \BM{h})=\BM{u}^\flat\wedge\nabla_u\BM{h}=
\BM{u}^\flat \wedge \BM{u}^\flat \wedge \BM{\xi}^\flat=0\,.
\ee
Thus for a geodesic motion, $\nabla_{u}\BM{u}=0$, the 2-form $\BM{F}$
is parallel propagated along the geodesic \cite{PKVK}.

The 2-form $\BM{F}$ has its own Darboux basis, which is called {\em
comoving}.  For any geodesic the comoving basis is determined
along its trajectory. Since $\BM{F}$ is parallel propagated, its
eigenvalues and its Darboux subspaces, which are called the {\em
eigenspaces} of  $\BM{F}$, are parallel-transported.   For a {\em
generic} timelike (or spacelike) geodesic the eigenspaces of $\BM{F}$
are at most 2-dimensional. In fact, the eigenspaces with non-zero
eigenvalues are 2-dimensional,  and the zero-value eigenspace is
1-dimensional for odd number  of spacetime dimensions and
2-dimensional for even. So, the comoving basis is defined up to
rotations  in each of the 2D eigenspaces. The  {\em
parallel-propagated} basis is a special comoving basis.  It can be
found by solving a set of the first order ordinary differential
equations for the angles of rotation in the 2D eigenspaces. It is
possible to show that these ordinary differential equations can be
solved by means of separation of variables \cite{CFK}. A modification
of this procedure can be used to construct parallel propagated frames
along {\em null geodesics} \cite{KFKC}.

\section{Degenerate case}

Consider a closed rank-$2$ CKY tensor $\BM{h}$ and denote by 
$x_{\mu}$ ($\mu=1,\cdots, n$)  and $\xi_i$  ($i=1,\cdots, N$) its
non-constant eigenvalues  and the non-zero constant ones,
respectively. Suppose the eigenvalues of the ``square of the CKY
tensor" $H=(H^a{}_b) = ( - h^{a}{}_c h^c{}_b )$  have the following
multiplicities:
\begin{equation}
\{
\underbrace{x_1^2, \dotsm,  x_1^2}_{2\ell_1},
\dotsm, 
\underbrace{x_n^2, \dotsm, x_n^2}_{2\ell_n}, 
\underbrace{\xi_1^2, \dotsm, \xi_1^2}_{2m_1},
\dotsm,
\underbrace{\xi_N^2, \dotsm, \xi_N^2}_{2m_N}, 
\underbrace{0,\dotsc, 0}_{K} \},  
\end{equation}
where $D=2 (|\ell| + |m| )+ K$.
Here $|\ell| = \sum_{\mu=1}^n \ell_{\mu}$ and $|m|=\sum_{i=1}^N m_i$. 
In the previous consideration we assumed that the principal CKY tensor
$\BM{h}$ is non-degenerate, that is 
\be
\ell_1=\ldots =\ell_n=1\hh
m_1=\ldots =m_N=0\hh
K=\ve\, .
\ee
If these conditions are violated we call a principal CKY tensor {\em
degenerate}. Recently there was obtained a general canonical form of
a \st \ which admits a degenerate principal CKY tensor
\cite{hoy1,hoy3}. These papers also contain a {\em generalized
Kerr-NUT-(A)dS metrics} which are solutions of the higher dimensional
Einstein equations with a cosmological constant with a arbitrary
degenerate principal CKY tensor. It is interesting to notice that a
4D Taub-NUT metric belongs to this class of the solutions.  An example
of a generalized Kerr-NUT-(A)dS with a degenerate principal CKY
tensor is a special subclass of Kerr-(A)dS solutions \cite{GLPP1,GLPP2}.
In a general case these solutions contain $[(D-1)/2]$ different
angular momenta. In  the case when some of them are equal, the
principal CKY tensor has constant eigenvalues \cite{TENSOR}.

At the moment not so much is known about properties and
interpretation of the generalized Kerr-NUT-(A)dS solutions in higher
dimensions. 

\section{Concluding remarks}
  
During past 2-3 years a lot of important and interesting results
concerning higher dimensional black hole solutions with the spherical
topology of the horizon were obtained. It was discovered that a key
role in this study is played by the principal CKY tensor. The very
existence of this tensor in a \st \ restricts the form of the metric.
In the non-degenerate case this canonical metric obeying the Einstein
equations coincides with  the Kerr-NUT-(A)dS solution. The class of
metrics admitting a principal CKY tensor allows separation of
variables in the Hamilton-Jacobi, Klein-Gordon, Dirac, and tensor
perturbation. It is still an open question whether other massless
fields equations, e.g. the Maxwell field, allow separation of
variables in the (generalized) Kerr-NUT-(A)dS \st. Another
interesting question concerns a possibility to separate variables in
the higher dimensional spacetimes in the absence of the principal CKY
tensor. It may happen when such a space admits a sufficient number
of commuting Killing vectors and Killing tensors \cite{BF}.
Especially, this problem of separation of variables is important for
study black ring and black saturn solutions\footnote{See e.g.
discussion of the integrability problem for geodesic motion near
black rings in \cite{Hos}}. Separation of variables for the
gravitational perturbations is important for the analysis of the
stability of these higher dimensional solutions. Another interesting
direction of research is study of the hidden symmetries in
supergravity black holes (see e.g. \cite{Chow}) and, more generally,
study of the relation between hidden symmetries and supersymmetry.

\medskip

\section*{Acknowledgements}

The author is grateful to the Natural Sciences and Engineering
Research Council of Canada and the Killam Trust for partial support.

\end{document}